%% file: alya-vector-size.tex
\documentclass[review]{elsarticle}

\input{00-preamble.tex}

\journal{Computers \& Fluids}

\begin{document}

\begin{frontmatter}

\title{A portable coding strategy to exploit vectorization on combustion simulations}
\author[bscmainaddress]{Fabio Banchelli}

\author[bscmainaddress]{Guillermo Oyarzun\corref{mycorrespondingauthor}}
\cortext[mycorrespondingauthor]{Corresponding author}
\ead{guillermo.oyarzun@bsc.es}

\author[bscmainaddress]{Marta Garcia-Gasulla}
\author[bscmainaddress]{Filippo Mantovani}
\author[bscmainaddress]{Ambrus Both}
\author[bscmainaddress]{Guillaume Houzeaux}
\author[bscmainaddress]{Daniel Mira}

\address[bscmainaddress]{Barcelona Supercomputing Center, Plaza Eusebi Guell, 1-3, 08034 Barcelona (Spain)}

\input{01-abstract.tex}


\end{frontmatter}


\input{10-intro.tex}

\input{20-app-context.tex}

\input{30-tech-context.tex}
\input{31-environment.tex}

\input{32-methodology.tex}

\input{40-analysis.tex}
\input{41-time.tex}
\input{42-instr-mix.tex}
\input{43-cache.tex}
\input{44-overall-performance.tex}
\input{50-conclusiones.tex}

\input{90-acks.tex}

\bibliography{alya-vector-size}

\end{document}

%% file: 00-preamble.tex
\usepackage[utf8]{inputenc}
\usepackage[T1]{fontenc}
\usepackage{amsmath,amssymb,amsfonts,mathrsfs}
\usepackage{algorithmic}
\usepackage{graphicx}
\usepackage{textcomp}
\usepackage[table,dvipsnames]{xcolor}
\usepackage{booktabs} 
\usepackage{multirow}
\usepackage{listings}
\def\BibTeX{{\rm B\kern-.05em{\sc i\kern-.025em b}\kern-.08em
    T\kern-.1667em\lower.7ex\hbox{E}\kern-.125emX}}

\usepackage{xspace}
\newcommand{\ie}{i.e.,\xspace}
\newcommand{\eg}{e.g.,\xspace}

\newcommand{\mn}{MareNostrum~4\xspace}

\newcommand{\skylake}{Skylake\xspace}

\newcommand{\vs}{\texttt{VECTOR\_SIZE}\xspace}
\xspaceaddexceptions{=}

\usepackage{listings} 
\definecolor{col1}{HTML}{1E88E5}
\definecolor{col2}{HTML}{D81B60}
\definecolor{col3}{HTML}{43A047}
\definecolor{col4}{HTML}{F4511E}
\definecolor{col5}{HTML}{205B74}

\usepackage{pifont}

\usepackage[colorlinks=true, allcolors=blue, bookmarks=false]{hyperref}

\RequirePackage[normalem]{ulem}
\RequirePackage{color}\definecolor{RED}{rgb}{1,0,0}\definecolor{BLUE}{rgb}{0,0,1}

\providecommand{\DIFdel}[1]{{\protect\color{RED}\sout{#1}}}
\providecommand{\DIFdel}[1]{}

\newcommand{\wt}[1]{\widetilde{#1}}

\newcommand{\vect}[1]{\boldsymbol{#1}}
\newcommand{\uV}{\vect{u}}
\newcommand{\tenso}[1]{\boldsymbol{#1}}

%% file: 01-abstract.tex
\begin{abstract}

The complexity of combustion simulations demands the latest high-performance computing tools to accelerate its time-to-solution results.
A current trend on HPC systems is the utilization of CPUs with SIMD or vector extensions to exploit
data parallelism.
Our work proposes a strategy to improve the automatic vectorization of finite-element-based scientific codes. 
The approach applies a parametric configuration to the data structures to help the compiler detect the block of codes that can take advantage of vector computation while maintaining the code portable.
A detailed analysis of the computational impact of this methodology on the different stages of a CFD solver is studied on the PRECCINSTA burner simulation.
Our parametric implementation has proven to help the compiler generate more vector instructions in the assembly operation: this results in a reduction of up to $9.39\times$ of the total executed instruction maintaining constant the Instructions Per Cycle and the CPU frequency.
The proposed strategy improves the performance of the CFD case under study up to $4.67\times$ on the \mn supercomputer.

\end{abstract}

%% file: 10-intro.tex
\section{Introduction and related work}\label{secIntro}


The decarbonization of the transportation sector is one of the fields with high
strategic importance for our society~\cite{UN2015,EU2015}.
Implementing new greener fuels in real combustion systems demands advanced combustion simulations, as their physical
 and chemical properties are expected to be significantly different from those of conventional transportation fuels~\cite{Zhang2016}. 
In such complex simulations,  the investigation of more accurate and efficient numerical algorithms is of key importance to increase
 the accuracy and reduce the time-to-solution.
The difficulty relies on the constant evolution of the High-Performance Computing (HPC) systems. Consequently, scientific software requires periodic updates to exploit the new features and run efficiently on those systems.

On modern CPUs, the use of vector or Single Instruction Multiple Data (SIMD) extensions is
becoming more and more relevant. 
%
Beside the AVX-512 SIMD extension by Intel, we detect appearing on the market the first CPU implementing the Arm SVE extension (Fujitsu A64FX, ranked first in the Top500) and the NEC SX-Aurora vector engine, a discrete accelerator leveraging vector CPUs able to operate with registers of up to 256 double precision elements.
On top of this market movements, we can not ignore the RISC-V architecture which recently ratified v1.0 of the V-extension, boosting vector computation from the academic world and the open-source community.

The efficient use of vector units within CPUs relies on auto-vectorization by the compiler and often requires to adapt or rewrite classical algorithms to exploit their full computing power~\cite{pabloFFT}.
Large-scale CFD codes are generally dominated by two operations: the linear solver and the matrix assembly. The first can be considered a black-box component
 that receives a matrix and a right-hand-side as an input and returns a solution
vector~\cite{Oyarzun2013}. The solver is composed of algebraic operations that can exploit
 vectorization by using specific libraries~\cite{eigen}. This strategy allows to port a part of large
 scientific codes to vector accelerators in a relatively smooth way~\cite{CODA}. 
 Regarding the matrix assembly, the algorithm for unstructured meshes depends on the 
discretization method, where finite volume (FV) or finite elements (FE) are the most common strategies.
 Obtaining gains from vectorization in FV assembly requires introducing changes that have
 proved not practical on large-scale combustion codes~\cite{YALES2, OpenFoam}. On the contrary, the FE assembly
 is constituted by matrix-like structures with the potential application of
SIMD-friendly functions~\cite{SIMD}.
Our work is implemented on Alya~\cite{Borrell2020}, a large-scale computational mechanics code (FE-based) that
 is one of the thirteen Unified European  Applications Benchmark Suite codes. 
We propose and analyze a parametric configuration to its data structure, allowing the compiler to enable auto-vectorization.
We evaluate the proposed implementation on a state-of-the-art supercomputer, \mn, powered by Intel Skylake CPUs.
We show that Alya takes advantage of AVX-512 SIMD units present in the Skylake CPUs while keeping the code portable.
The strategy is extensible to any other FE-based code.

The main contributions of this paper are:
{\em i)}
we propose a parametric configuration of the data structure for a complex fluid-dynamic code;
{\em ii)}
we measure and explain the impact of the proposed configuration from a computational point of view;
{\em iii)}
we quantify the overall performance gain on a state-of-the-art HPC supercomputer.

The remaining part of the paper is structured as follows:
Section~\ref{secAppContext} summarizes the computational fluid-dynamics problem solved with Alya;
Section~\ref{secTechContext} briefly presents the technological context of the study performed in this paper, including details of the hardware and software configurations.
Section~\ref{secPerf} analyzes the optimizations applied to Alya in terms of execution time, instruction mix and cache effects to quantify the overall performance gain.
Section~\ref{secConclusions} closes the paper with general remarks and conclusions.

%% file: 20-app-context.tex
\section{Application context}\label{secAppContext}

\subsection{Governing equations}

The governing equations describing the reacting flow field in the turbulent premixed flame correspond to the
 low-Mach number approximation of the Navier-Stokes equations with the energy equation represented by the
 total enthalpy. The combustion process is assumed to take place in the flamelet regime and the flamelet 
database is based on the tabulation of a laminar premixed flamelet at constant
equivalence ratio that uses 
the chemistry from the San Diego mechanism~\cite{SanDiego}. A Favre-filtered description of the governing 
equations is followed to avoid modelling of terms including density fluctuations~\cite{Poinsot}.
The governing equations are given by:
 
\begin{equation}
\label{e:con}
\frac{\partial \overline{\rho}}{\partial t} + \nabla \cdot \left( \overline{\rho} \widetilde{\mathbf{u}} \right) = 0
\end{equation}
\begin{equation}
\label{e:m}
\frac{\partial\left( \overline{\rho} \widetilde{\mathbf{u}}\right)}{\partial t} + \nabla \cdot \left(\overline{\rho} \widetilde{\mathbf{u}}\widetilde{\mathbf{u}}\right) = -\nabla \overline{p} + \nabla \cdot \left [\overline{\rho} (\overline{\nu}+\nu_t) \left( 2\tenso{S} - \frac{2}{3} \left(\nabla\wt{\uV}\right) \tenso{I} \right) \right ]
\end{equation}
\begin{equation}
\label{e:h}
\frac{\partial\left(\overline{\rho} \widetilde{h}\right)}{\partial t} + \nabla \cdot \left( \overline{\rho} \widetilde{\mathbf{u}} \widetilde{h} \right) = \nabla \cdot \left[  \overline{\rho} \left( \overline{D} + \frac{\nu_t}{Pr_t} \right) \nabla \widetilde{h} \right]
\end{equation}
where $\overline{\rho}$, $t$, $\widetilde{\rm{u}}$,  $\overline{p}$, $\overline{\nu}$, $\widetilde{h}$ and
 $\overline{D}$ represent the density, time, velocity vector, pressure, kinematic viscosity, total 
enthalpy and thermal diffusion coefficient respectively. 
Heating due to viscous forces is neglected in the enthalpy equation 
and the unresolved heat flux is modelled using a gradient diffusion approach~\cite{Daniel_IJHE_2014}. The 
formulation is closed by an appropriate expression for the subgrid-scale or eddy-viscosity $\nu_{t}$ that 
in this study is defined by the closured proposed by Vreman \cite{Vreman2004} with a model constant 
of $c_s=0.1$.
The viscous stress tensor is defined based on Stokes' assumption and the turbulence contribution is determined 
by the use of the Boussinesq approximation~\cite{Poinsot}, in which
 $\tenso{S} = \frac{1}{2} \left[ \nabla \wt{\uV} + \left( \nabla \wt{\uV} \right)^T \right]$
 and $\tenso{I}$ are the strain and the identity tensor respectively. A unity Lewis number assumption has 
been made to simplify the multicomponent transport in the governing equations.
Turbulent Schmidt and Prandtl numbers are both set constant with value of $0.7$.

For the present combustion model, a controlling variable based on a reactive scalar is used to couple the
 chemical states with the fluid flow. 
This controlling variable can be understood as a progress variable $Y_c$ 
that is used to describe the thermochemical state from an unreacted mixture to
a fully reacted mixture.
For numerical reasons~\cite{Domingo_CF_2005}, a scaled progress variable $c$ is defined as:
\begin{equation}
c = \frac{Y_c - Y_c^0}{Y_c^{eq} - Y_c^0}
\end{equation}
where $Y_c^0$ and $Y_c^{eq}$ are the values of the progress variable of the unreacted mixture and at
 equilibrium conditions respectively. 
Considering the application of this flamelet combustion model to premixed combustion in LES, the 
subscale effects need to be addressed. 
The tabulated properties $\psi$ are integrated with a presumed-shape probability density function (PDF) that
is constructed from the filtered progress variable $\widetilde{c}$ and the subgrid
 variance $\widetilde{c''^2}=\widetilde{cc}-\widetilde{c}\widetilde{c}$ using a
 $\beta$-function~\cite{Domingo_CF_2005}. A closure for the subgrid scale variance is provided by 
the solution of the transport of $\widetilde{c''^2}$ following Domingo et al. (2005)~\cite{Domingo_CF_2005}. 

The chemical state of the perfectly premixed flame in the LES framework is ultimately described by 
the two controlling variables: $\widetilde{c}$ and $\widetilde{c''^2}$, so the governing equations 
describing the chemical evolution of the flame are given by:

\begin{equation}
\label{e:c}
\frac{\partial\left( \overline{\rho} \widetilde{c} \right)}{\partial t} + 
 \nabla \cdot \left( \overline{\rho} \widetilde{\mathbf{u}} \widetilde{c} \right) =
 \nabla \cdot \left[ \overline{\rho} \left( \overline{D} + \frac{\nu_t}{Sc_t} \right) \nabla \widetilde{c} \right] + \overline{\dot \omega}_{c}
\end{equation}
\begin{align}
\label{e:cv}
\frac{\partial\left( \overline{\rho} \widetilde{c''^2} \right)}{\partial t} + 
 \nabla \cdot \left( \overline{\rho} \widetilde{\mathbf{u}} \widetilde{c''^2} \right) = &
 \nabla \cdot \left[ \overline{\rho} \left( \widetilde{D} + \frac{\nu_t}{Sc_t} \right) \nabla \widetilde{c''^2} \right]  \\
 & +2 \overline{\rho}  \widetilde{D} \left |\nabla \widetilde{c} \right|^2 \\
&+ 2 \left (\overline{c \dot{\omega}_{c}} - \widetilde{c} \overline{\dot{\omega}_{c}}\right ) \\
&- \overline{\rho}\widetilde{\chi}_{c}  
\end{align}
where $\widetilde{\chi}_{c}$ represents the scalar 
dissipation rate and $\overline{\dot \omega}_{c}$ is the filtered source term
of the progress variable. The scalar dissipation rate is composed by the resolved and unresolved parts, 
which are given by:

\begin{equation*}
\widetilde{\chi}_{c} = 2\widetilde{D}\left | \nabla \widetilde{c} \right |^2 + \chi_{c}^{sgs} = 2\widetilde{D}\left | \nabla \widetilde{c} \right |^2 + \frac{C_d}{\tau_t}\widetilde{c''^2}
\end{equation*}
where $\tau_t$ is a turbulent time scale, which is  obtained following Ventosa et al.~\cite{Ventosa_FTaC_2017}. 
\subsection{About Alya}
\label{sec:about_alya}
The governing equations (\ref{e:con}), (\ref{e:m}), (\ref{e:h}), (\ref{e:c}), and (\ref{e:cv}) are solved by means of a 
low-dissipation finite-element method implemented into the code Alya~\cite{Vazquez_JCS_2016}. The code Alya has been used 
to resolve turbulent reacting flows in premixed and partially premixed 
conditions~\cite{Govert_AE_2015,Govert_FTaC_2018,mira2020numerical,
Both_CAF_2020}. The use case uses a perfectly 
premixed model with pressumed-shape PDF to account for turbulent-chemistry interactions. The convective term is 
discretized using an extension to variable density flows of the scheme recently proposed by Charnyi et al. \cite{Charnyi2017}, which 
conserves linear/angular momentum and kinetic energy at the discrete level. Second-order spatial discretizations are used.  In order 
to use equal-order elements, numerical dissipation is introduced only for the pressure stabilization via a fractional 
step scheme~\cite{Codina01}. The set of equations is integrated in time using a third-order Runge-Kutta explicit method 
combined with an eigenvalue-based time-step estimator~\cite{Trias2011}. This approach has been shown to be significantly less 
dissipative than traditional stabilized FEM approach~\cite{Lehmkuhl2019}. Scalar equations are solved by a third-order 
explicit Runge-Kutta scheme combined with the ASGS stabilization method~\cite{Both_CAF_2020}.

\begin{figure}[htbp]
   \centering
   \includegraphics[width=\columnwidth]{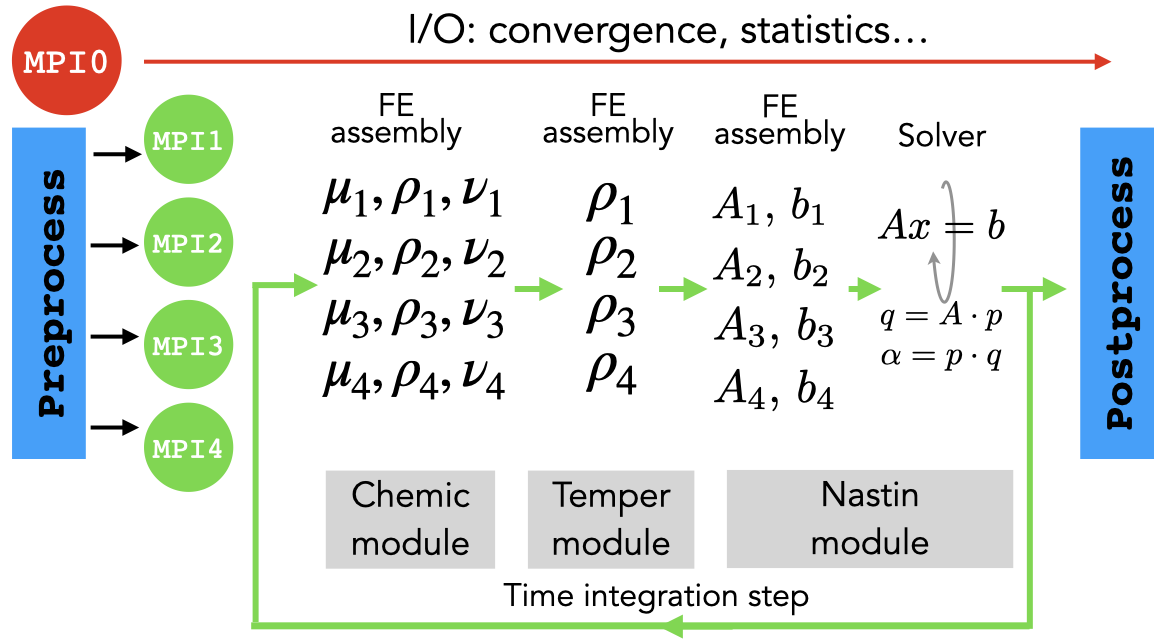}
   \caption{Workflow of Alya for a combustion simulation.}
   \label{fig:workflow}
 \end{figure}

Alya is a modular scientific code in which each module handles a set of equations. The momentum equations are solved by 
the \textit{nastin} module that applies the fractional step method. The algorithm consists of two procedures: 
the application of the third-order Runge-Kutta for the discretization of the convection-diffusion equation and the solution of the Poisson 
equation that imposes the mass conservation constraint. The Runge-Kutta is an iterative method that assembles the laplacian matrix 
and a right-hand side that forms the Poisson system. This linear system is solved utilizing an iterative Krylov method. 
The \textit{temper} module is responsible for calculating the contribution of the energy equation, and the \textit{chemic} module 
manages the chemical transport equations. All modules define a unique thermo-chemical state of the simulated gas flow. 
These modules are based on an explicit scheme that assembles the arrays related to thermal conductivity, viscosity, and variable density. Latter creates a strong coupling between \textit{nastin} and the other modules, since the temporal 
derivative of density appears as a source term in the continuity equation used in the fractional step method. A general view 
of the Alya's workflow is shown in Figure~\ref{fig:workflow}. The simulation of combustion phenomena requires of millions of time 
integration steps to attain meaningful results. The operations within the time-integration step become the dominant ones.

The assembly procedure is present in all the modules since it is an essential part of the finite element method. 
The assembly is applied on each element locally, not requiring MPI communications. The primary operations within
 the time integration step are the element assembly and the algebraic solver. Table~\ref{tab:matass} summarizes 
the relative weight of the assembly within the time-integration step for a
combustion simulation using the flamelet model.

\begin{table}[htbp]
\caption{Relative weight of the main operations within the time integration step}
\begin{center}
\begin{tabular}{|c|c|c|c|c|}
\hline
\textbf{} & \multicolumn{4}{|c|}{\textbf{Numerical Equations}} \\
\cline{2-5} 
\textbf{Module} & \textbf{\textit{nastin}}& \textbf{\textit{temper}}& \textbf{\textit{chemic}}& \textbf{\textit{Total}} \\
\hline
\textbf{Element Assembly} & $43.14\%$ & $12.66\%$ & $39.57\%$ & $95.37\%$  \\
\textbf{Algebraic Solver} & $4.47\%$ & $-$ &  $-$  & $4.47\%$
\\
\textbf{Others} & $0.04\%$ & $0.03\%$ & $ 0.09\%$  & $0.16\%$
\\ \hline
\textbf{Total} & \textbf{$47.65\%$} & \textbf{$12.69\%$} & \textbf{$39.66\%$} & \textbf{$100\%$}  \\
\hline
\end{tabular}
\label{tab:matass}
\end{center}
\end{table}

The number of iterations in the element assembly depends on two variables: the
number of gauss points (\textit{ngauss}) and the number of integration nodes (\textit{nnodes}).  The values
 of \textit{ngauss} and \textit{nnodes} varies according to the shape of the element (cell of the mesh) 
and the discretization order.  Using a first-order discretization on linear elements produces an equal 
number of integration nodes and gauss points.  For instance, in tetrahedrons, prisms, and hexahedrons, 
the number of integration points is four, six, and eight, respectively.

The elemental matrix (Ae) is calculated using the shape functions (N) and the Jacobian element matrix (Jac). 
The traditional approach consists of assembling one after the other as shown in Algorithm~\ref{lst:simple}. 
The number of integration nodes and gauss points might vary from one element to another, thus preventing the 
automatic vectorization by the compiler of the inner loop.
At the end of the assembly phase, a reduction along the gauss points calculates the contribution of each element.

\begin{lstlisting}[basicstyle={\small}, caption={Element assembly},label=lst:simple]
do ig = 1,ngauss
  do jn = 1,nnodes
    do in = 1,nnodes
      Ae(in,jn) = Ae(in,jn) + Jac(ig) * N(in,ig) * N(jn,ig)
    end do
  end do
end do
\end{lstlisting}

\subsection{Implementation details}
\label{sec:impl}
One of the most important constraints when developing a scientific code adopted by a large community is that the code must be understandable, 
easy to maintain, and portable. One way to comply with these requirements is to limit (and if possible avoid) compiler-specific and architecture-specific implementations. Ideally one would like to squeeze the maximum performance avoiding coding styles that over-specialize the implementation of the scientific application. In the case of data parallelism, the hope is that the compiler detects 
the vectorizable zones (\eg loops) without the need of using specific SIMD code / assembly.
The software design approach presented in Algorithm~\ref{lst:simple} fails to unlock 
the potential vectorization and data reuse that exists within the assembly algorithm.
To leverage the performance boost delivered by SIMD / vector units of modern CPUs, the following set of preprocessing functions are proposed:

\begin{itemize}
    \item \textbf{Grouping}: The elements are organized in groups. Each group contains elements 
    with the same geometrical form. Then, the elemental assembly within a group has the same number of 
    gauss points and integration points. This reorganization creates a regularity in the operations 
    performed within elements of the same group. The goal is to unlock the potential vector operations. 
    \item \textbf{Renumbering:} The elements within a group are renumbered using a Cuthill-Mkee algorithm.
    The idea is to minimize the cache misses by reducing the bandwidth of the connectivity matrix. 
    \item \textbf{Packing:} Each group is divided into packs of \texttt{VECTOR\_SIZE} elements. The 
    definition of the \texttt{VECTOR\_SIZE} takes place at compilation time. The elemental 
    multi-dimensional arrays involved in the assembly operation incorporate a new dimension of 
    size \texttt{VECTOR\_SIZE}.  In Fortran, the extra dimension is added at the first position of the 
    arrays since  the memory distribution follows a column-major order.  The idea is to perform the 
    assembly operations to all the elements of a pack at once. The column-major order and the extra 
    dimension help the compiler to generate vector operations. 
    \item \textbf{Padding zeros:} If the number of elements within a group is not divisible by 
    the \texttt{VECTOR\_SIZE}, zeros are padded in the elemental arrays to maintain the 
    regularity of the pack.
\end{itemize}

In our case, the proposed functions have been integrated into Alya. The
implementation of those depends heavily on the specific scientific code: parallelization strategy,
 internal data structures, and programming language.
The preprocessing functions are called only once at the beginning of the simulation.
 An Alya execution consists of thousands or millions of iterations, making the cost of the
 preprocessing stage negligible; therefore, these functions are not further studied in this manuscript. 

Additionally, a data structure stores the relevant information of each group, \ie number of 
gauss points, integration points, and element ids. Such data structure works as an index that 
allows jumping between packs, gathering the global data, and scattering the outcome results. 
For instance, the assembly operation using \texttt{VECTOR\_SIZE} is shown at Algorithm~\ref{lst:vec}. 
Using the subscript-triplet notation of Fortran (\texttt{1:VECTOR\_SIZE}) provides extra 
information to the compiler for unlocking the vectorization.  Note that performing the operations 
in packs also exposes the temporal locality of the reusable arrays as the shape
functions. Moreover, using \texttt{VECTOR\_SIZE=1} is equivalent to the original code in terms of memory accesses and performance.
Our study focuses on the impact of this parametric variable on enabling the SIMD instructions and improving memory accesses.

\begin{lstlisting}[basicstyle={\small}, caption={Matrix element assembly using VECTOR\_SIZE},label=lst:vec]
do ig = 1,ngauss
  do jn = 1,nnodes
    do in = 1,nnodes
      Ae(1:VECTOR_SIZE,in,jn) = Ae(1:VECTOR_SIZE,in,jn) + Jac(1:VECTOR_SIZE,ig) * N(in,ig) * N(jn,ig)
    end do
  end do
end do
\end{lstlisting}

\subsection{Application context: The PRECCINSTA burner}
\label{sec:preccinsta}
\begin{figure}[h!]
    \centering
    \includegraphics[width=\linewidth]{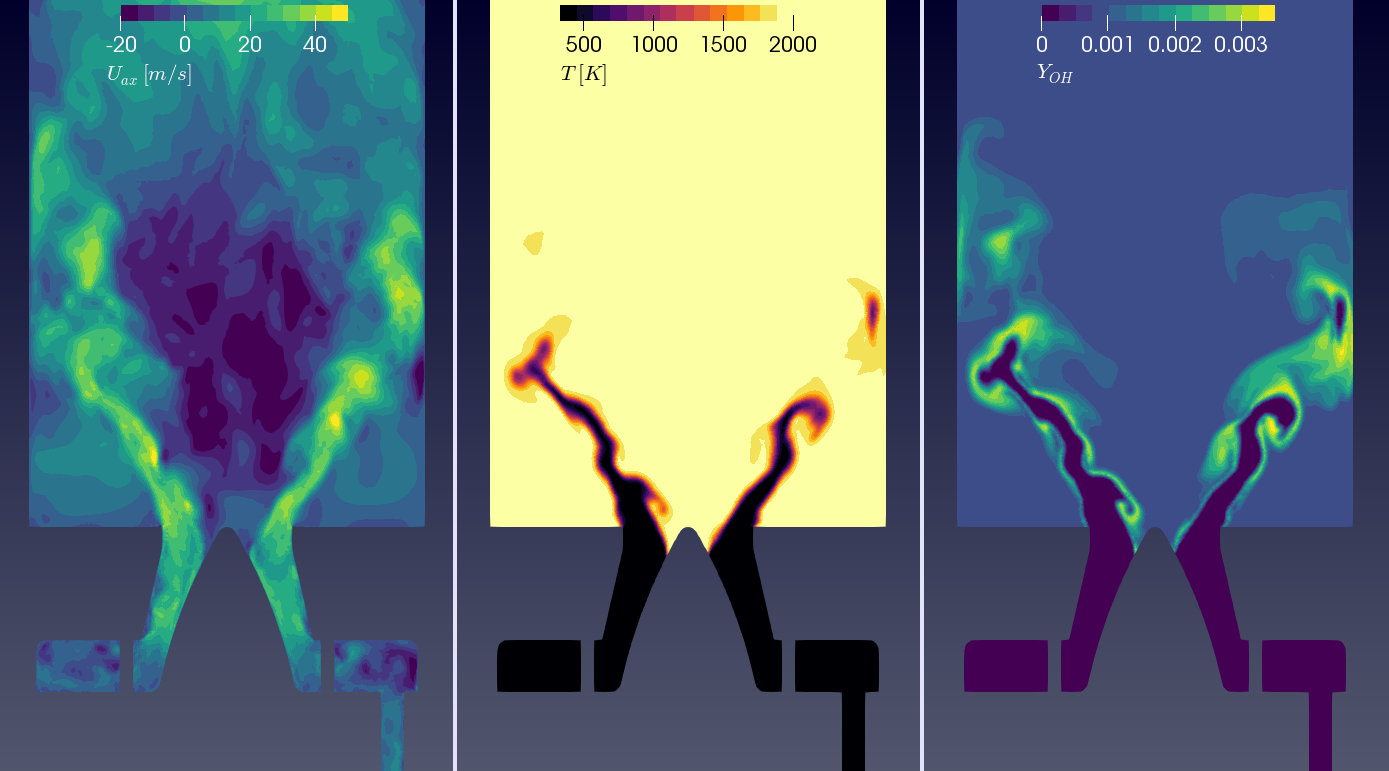}
    \caption{Instantaneous contour plots of the LES results of the PRECCINSTA burner. Left: axial velocity, middle: temperature, right: hydroxyl radical mass fraction. }
    \label{fig:preccinsta_contours}
\end{figure}

The use case is a premixed swirl-stabilized flame of a gas turbine model combustor, also known as the
PRECCINSTA burner ~\cite{Meier2007}. This is a traditional benchmark for
combustion simulations that permits to evaluate the main operations involved in
production runs. Alya is set up to run a large-eddy simulation using the flamelet
 method for calculating the tabulated chemical transport. The burner operates at atmospheric pressure and ambient
temperature with an equivalence ratio of $\phi = 0.67$. The domain
discretization consists of an unstructured mesh of 16 million elements with different shapes: tetrahedrons, pentahedrons, and pyramids.
Most of the elements are tetrahedrons ($76.4$\%), followed by pentahedrons(
$23.5$\%), while the rest are pyramids ($0.1$\%) used to smooth the transition between the other two.
 After the preprocessing stage, twenty-six elements with empty entries are
padded to maintain regularity for a \texttt{VECTOR\_SIZE=32}; this
amount of padded elements is insignificant compared to the 16 million element mesh.
The loop size in Algorithm~\ref{lst:vec} takes a different number of gauss points depending on the element shape. The tetrahedrons, pyramids, and pentahedrons require four, five, and six gauss points, respectively. 
 Sample results of the Large Eddy Simulation (LES) fields are shown in Figure~\ref{fig:preccinsta_contours} for the axial velocity, temperature, and hydroxyl (OH) radical.
 Details of the analysis and validation of this case can be found in Govert et al.\cite{Govert_FTaC_2018}, and Both et al.~\cite{Both_CAF_2020}.

%% file: 30-tech-context.tex
\section{Technological context}\label{secTechContext}

In this section we describe the hardware and software as well as the methodology and metrics that we employ in our study.

%% file: 31-environment.tex
\subsection{Environment}\label{secEnv}

\mn is the flagship supercomputer at the Barcelona Supercomputing Center.
Table~\ref{tabHardware} summarizes the hardware characteristics of \mn.
The cluster is composed of~3456 compute nodes. Each node houses two Intel Xeon Platinum~8160 CPU with~24 cores running at~2.10 GHz.
The CPUs implement the Intel x86 Skylake microarchitecture and support SIMD instructions SSE, AVX, and AVX-512 which operate with registers of~128,~256, and~512 bits respectively.
For example, the AVX-512 registers can hold up to eight double precision floating point elements.

\begin{table}[htbp]
\centering
\caption{Hardware configuration of \mn} \label{tabHardware}
\rowcolors{2}{gray!15}{white}
\begin{tabular}{l|c}
System integrator                 & Lenovo            \\
Core architecture                 & Intel x86         \\
SIMD extensions                   & SSE, AVX, AVX-512 \\
CPU name                          & \skylake Platinum \\
Frequency {[}GHz{]}               & 2.10              \\
Sockets/node                      & 2                 \\
Core/node                         & 48                \\ \midrule

L1 cache size                     & 64 kB             \\
L2 cache size                     & 256 kB            \\
L3 cache size                     & 33 MB             \\
Memory/node {[}GB{]}              & 96                \\
Memory tech.                      & DDR4-2666         \\
Memory channels                   & 6 per socket      \\  
Peak memory bandwidth {[}GB/s{]}  & 256 GB/s          \\ \midrule

Number of nodes                   & 3456              \\
Interconnection                   & Intel OmniPath    \\  
Peak network bandwidth {[}GB/s{]} & 12.00             \\

\end{tabular}%
\end{table}

The approach we present in this paper is portable because it is independent from the architecture and the compiler. However, as we rely on the autovectorization capabilities of the compiler, the performance results can depend on the compiler used. 
We compiled Alya with three different compilers based on software availability, and having vendor specific compilers (Intel) and generic ones (GNU).
For both compilers we used the flags suggested by the application developers and the support team of the cluster on which we were running.
When compiling binaries with the Intel Compiler, we used the optimization flags \texttt{-xCORE-AVX512 -mtune=skylake};
whereas when compiling with the GNU compiler we used the flags \texttt{-O3 -march=skylake-avx512 -ffp-contract=fast -ffast-math}.
In addition to the optimization flags, the \texttt{-DVECTOR\_SIZE=<x>} defines, at compile time, the element packing as explained in Subsection~\ref{sec:impl}. Where the value of \texttt{<x>} is $\left\{1, 2, 4, 8, 16, 32, 64, 128, 256, 512 \right\} $.
As the parallel MPI performance is not part of the study, we use the same MPI library (\ie Intel MPI 2018.4) for all runs.

%% file: 32-methodology.tex
\subsection{Methodology}\label{secTestbed}

All experiments and results shown in this paper are obtained using a single node of the \mn cluster (48 cores) and launching 48 MPI ranks.
For this study, we follow a top-down approach, from more general metrics to more detailed ones. The goal is to understand the inherent behaviour of the different executions we are comparing.
We leveraged the hardware counters in \mn to gather information during the execution of Alya. To read these hardware counters we use PAPI~\cite{mucci1999papi,terpstra2010} combined with Extrae~\cite{extrae}.
Table~\ref{tabCounters} lists the counters we included in our study and a brief description of the events that they measure.
We instrumented the code to trigger an Extrae event that measures the hardware
counters at the start and the end of each phase in a time-integration step.

\begin{table}[htbp]
\caption{List of hardware counters} \label{tabCounters}
\rowcolors{2}{gray!15}{white}
\begin{tabular}{l|l}
\textbf{Name}                                & \textbf{Description}                                \\ \midrule
  \texttt{UNHALTED\_REFERENCE\_CYCLES}       & CPU cycles                                          \\
  \texttt{INST\_RETIRED}                     & Total number of executed instructions               \\
  \texttt{FP\_ARITH:SCALAR\_DOUBLE}          & Scalar floating point arithmetic instructions       \\
  \texttt{FP\_ARITH:128B\_PACKED\_DOUBLE}    & 128-bit SIMD floating point arithmetic instructions \\
  \texttt{FP\_ARITH:256B\_PACKED\_DOUBLE}    & 256-bit SIMD floating point arithmetic instructions \\
  \texttt{FP\_ARITH:512B\_PACKED\_DOUBLE}    & 512-bit SIMD floating point arithmetic instructions \\
  \texttt{BRANCH\_INSTRUCTIONS\_RETIRED}     & Branch instructions                                 \\
  \texttt{MEM\_UOPS\_RETIRED:ALL\_LOADS}     & Load micro-operations                               \\
  \texttt{MEM\_UOPS\_RETIRED:ALL\_STORES}    & Store micro-operations                              \\
  \texttt{MEM\_LOAD\_UOPS\_RETIRED:L1\_HIT}  & L1 cache hits produced by load operations           \\
  \texttt{MEM\_LOAD\_UOPS\_RETIRED:L1\_MISS} & L1 cache misses produced by load operations         \\
  \texttt{MEM\_LOAD\_UOPS\_RETIRED:L2\_HIT}  & L2 cache hits produced by load operations           \\
  \texttt{MEM\_LOAD\_UOPS\_RETIRED:L2\_MISS} & L2 cache misses produced by load operations         \\
  \texttt{MEM\_LOAD\_UOPS\_RETIRED:L3\_HIT}  & L3 cache hits produced by load operations           \\
  \texttt{MEM\_LOAD\_UOPS\_RETIRED:L3\_MISS} & L3 cache misses produced by load operations         \\
\end{tabular}%
\end{table}

When reading hardware counters, we run Alya with five time-integration steps per execution.
For each execution, we gathered the hardware counters information when the processes are performing useful computation (\ie not during MPI calls). 
\begin{figure}[htbp!]
  \centering
  \includegraphics[width=.95\columnwidth]{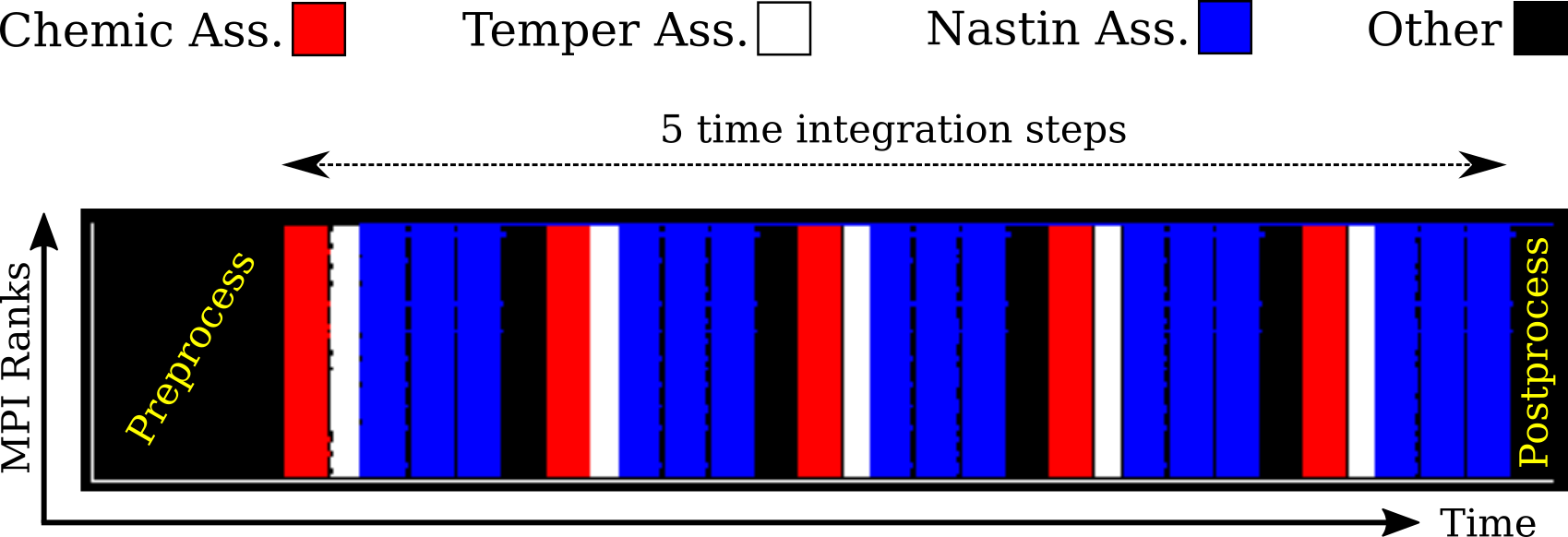}
  \caption{Example of trace obtained to measure hardware counters showing the different phases}
  \label{fig:trace5steps}
\end{figure}

In Figure~\ref{fig:trace5steps}, we show an example of a trace obtained to
measure the different hardware counters. On the $x$-axis we represent the time,
and in the $y$-axis the different MPI processes. The color shows the event added
with Extrae to delimiter the FE assembly of the different modules. If we compare
it with the workflow shown in Figure~\ref{fig:workflow} we can easily identify
the different phases: Preprocess, 5 time-integration steps and postprocess.

\begin{figure}[htbp!]
  \centering
  \includegraphics[width=.95\columnwidth]{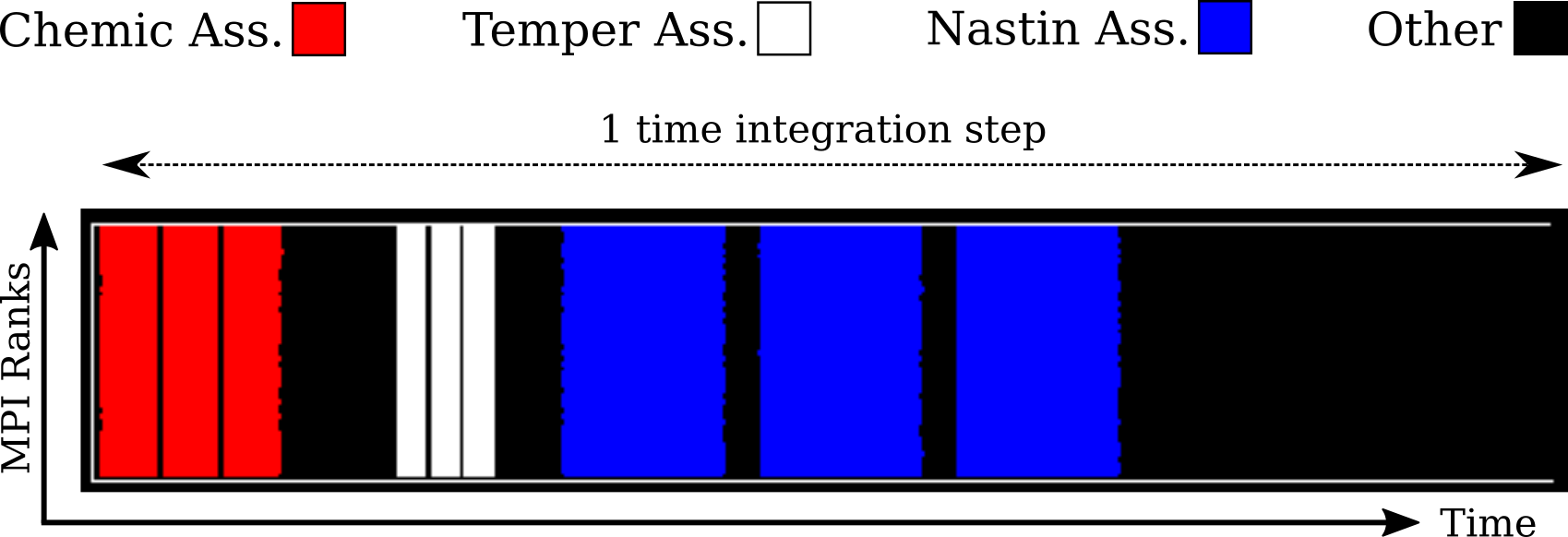}
  \caption{Example of trace obtained to measure hardware counters showing 1 time step}
  \label{fig:trace1step}
\end{figure}

In Figure~\ref{fig:trace1step}, we zoom-in focusing on one
time-integration step of the same trace. For each module execution, there are
three FE assembly calls due to the third-order Runge-Kutta used in the explicit
formulation of Alya (see Section~\ref{sec:about_alya}). 
The remainder corresponds to the algebraic solver that
is called only once per time-integration step.
We aggregate the values of the hardware counters obtained in all the regions with the same color.
This means that the measurements reported in Subsection~\ref{secInstrMix} and Subsection~\ref{secCache} represent the sum of all the events recorded by a given hardware counter across all processes in a given phase (\ie in a region with the same color).

%% file: 40-analysis.tex
\section{Performance analysis}
\label{secPerf}

In this section we study the performance impact of the coding changes explained in Section~\ref{sec:impl} in the environment and using the methodology explained in Section~\ref{secEnv}.
The study is guided by the \textit{computing performance equation} where the execution time $t$ of a program is computed as:
\begin{equation}
\label{perfEq}
t=\frac{I}{\mathscr{C} * f}
\end{equation}

where $\mathscr{C}$ stands for \textit{Instructions Per Cycle} and measures the efficiency of the processor in terms of how many instructions can be processed in one clock cycle. $I$ is the total number of instructions executed and $f$ the frequency at which the processor is working. 

The remaining part of this section analyzes each of the operands
of the \textit{computing performance equation} expressed in Equation~\ref{perfEq}: first we measure the elapsed time and then we study $f$ (the frequency), $I$ (the number and the types of instructions involved in the computation), and $\mathscr{C}$ (the IPC which we demonstrate is correlated with the cache reuse).

%% file: 41-time.tex
\subsection{Elapsed Time analysis}\label{secTime}

For the elapsed time analysis we use the elapsed time per phase, each value reported in this section is an average of~5 time steps. 
Alya follows an iterative pattern were each iteration performs the same computation. We verified that the variations of our measurements do not exceed 3$\%$. 5 time steps is a tradeoff between statistical significance, execution time of the tests and size of the traces collected.

\begin{figure}[htbp!]
  \centering
  \includegraphics[width=.95\columnwidth]{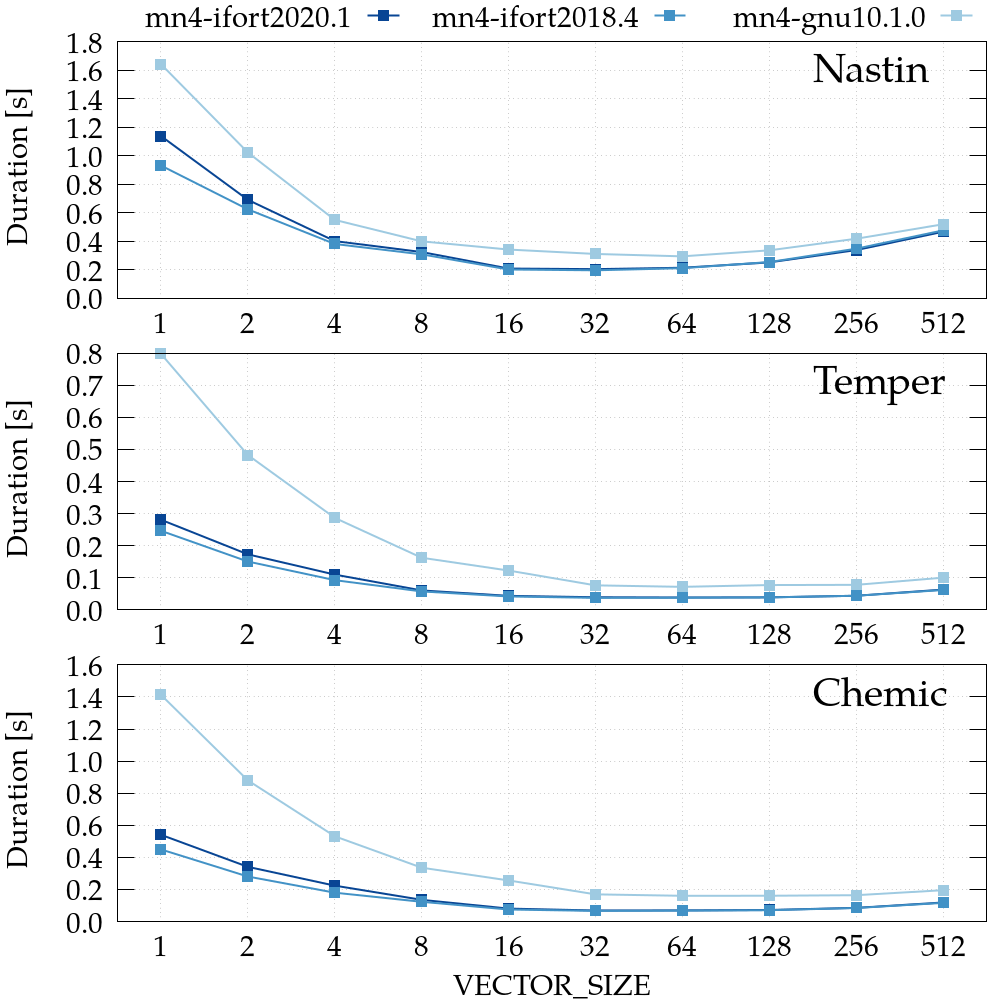}
  \caption{Elapsed time per phase and time step varying the vector size with different compilers}
  \label{figTimesteps}
\end{figure}

In Figure~\ref{figTimesteps} we show three plots corresponding to the three phases that we are studying, top to bottom: Nastin, Temper and Chemic. In the $y$-axis we can see the average elapsed time in a given phase and in the $x$-axis the value used for the \texttt{VECTOR\_SIZE}, each line corresponds to a different compiler.

Measurements of the master process (rank zero) are discarded. Thus, each point in Figure~\ref{figTimesteps} represents the elapsed time averaged across~47 processes, five time steps, and three assembly steps.
 
In Figure~\ref{figTimesteps} we observe that in all phases the GNU compiler obtains a worse performance than the different versions of the Intel compiler, this difference is more important for the Temper and Chemic phases than for Nastin. Also, for Temper and Chemic the difference is more important for low values of the \texttt{VECTOR\_SIZE}.

Comparing the different versions of the Intel compiler we also see a difference in performance. For \texttt{VECTOR\_SIZE} values below 16 Intel 2020 generates a code that performs better than the code generated with the 2018 version.

Increasing the \texttt{VECTOR\_SIZE} up to a value of 32 reduces the elapsed time for all compilers.
When using values of \texttt{VECTOR\_SIZE} greater than 32 there is a slight degradation of the performance for all the phases and all the compilers. However, the performance degradation is more important for the Nastin assembly.
To better explain these results, we look in more details at the hardware counters.

\subsection{Cycles and Frequency}\label{secCyc}

In this subsection we analyze the data obtained with the hardware counter \texttt{UNHALTED\_REFERENCE\_CYCLES}, that counts the total number of clock cycles. With this hardware counter we compute two metrics, the first one the total number of cycles used in a phase to do useful work, $C_{tot}$, and the second one the number of cycles per $\mu s$. The cycles per $\mu s$ can also be expressed as the measured frequency, $F$, and is computed as: $F = C_{tot} / T_{tot}$, where $T_{tot}$ is the total time spent in a phase while performing useful work.
 
\begin{figure}[htbp]
  \centering
  \includegraphics[width=.95\columnwidth]{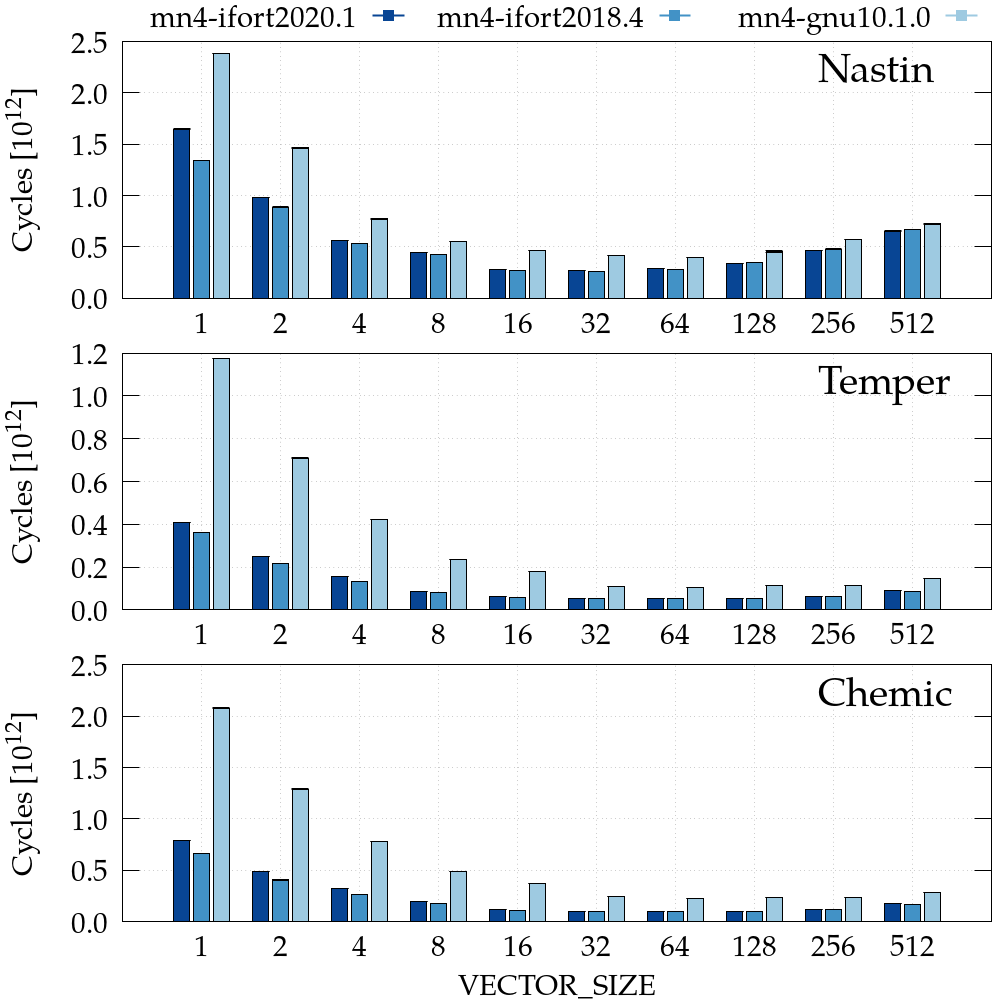}
  \caption{Cycles used to compute each phase varying the vector size with different compilers}
  \label{figCycles}
\end{figure}

In Figure~\ref{figCycles} we show the number of cycles used to compute each phase in the $y$-axis. In the $x$-axis we see the different values of the \texttt{VECTOR\_SIZE}. 
With a constant frequency and knowing that there is no communication during the measured time, the results in this plot should express the same trend as the one of the elapsed time depicted in Figure~\ref{figTimesteps}. 

\begin{figure}[htbp]
  \centering
  \includegraphics[width=.95\columnwidth]{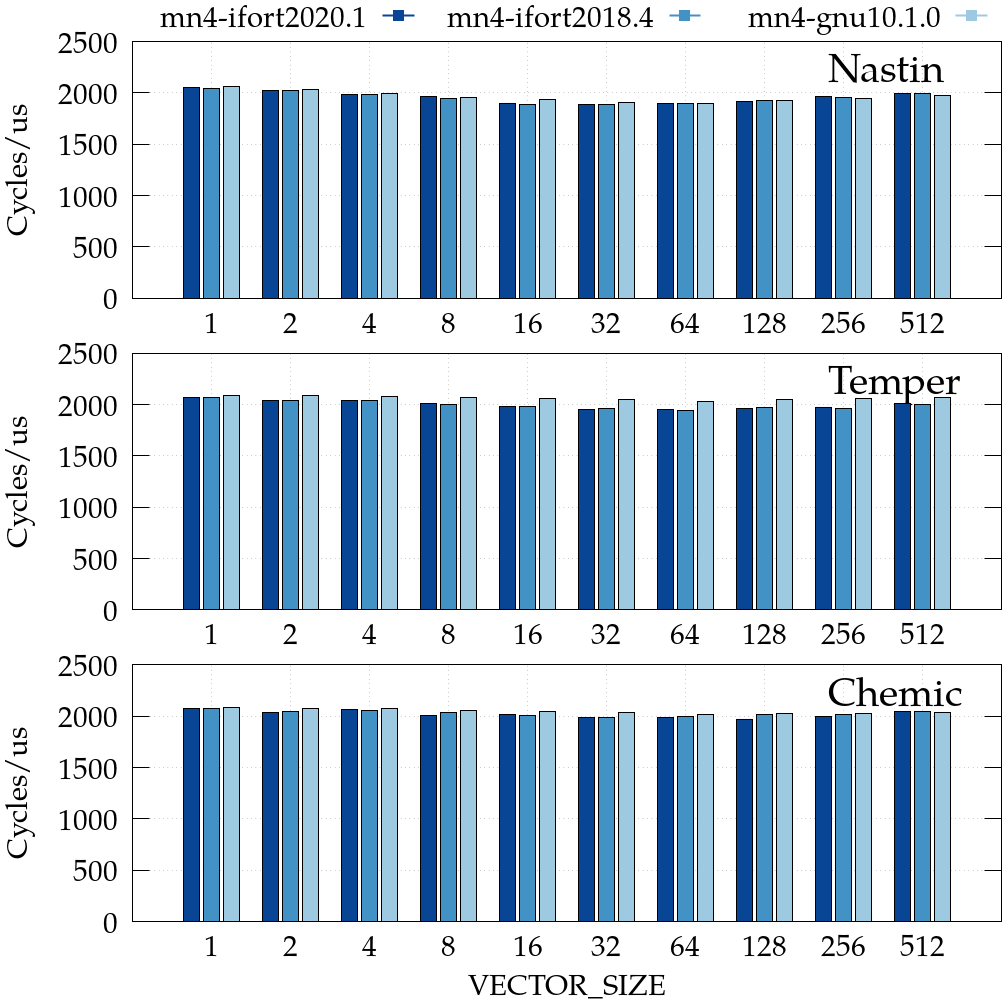}
  \caption{Frequency measured in each phase varying the vector size with different compilers}
  \label{figFrequency}
\end{figure}

In Figure~\ref{figFrequency} we show the frequency ($y$-axis) for the different executions and phases while changing the  \texttt{VECTOR\_SIZE} ($x$-axis). We can verify that there are no major changes in the frequency when varying the \texttt{VECTOR\_SIZE}, the phase nor the compiler.

It important to note that the operational frequency of the CPUs of \mn is set to 2.1~GHz, Dynamic Voltage and Frequency Scaling (DVFS) is disabled and we verified that there is no throttling of the frequency due to thermal protection (even when enabling AVX-512).

From this part of the study we can conclude that the differences that we observe in the elapsed time in Section~\ref{secTime} are not explained by a variation in the execution frequency.

%% file: 42-instr-mix.tex
\subsection{Instruction mix analysis}\label{secInstrMix}

\begin{figure}[htbp]
  \centering
  \includegraphics[width=.95\columnwidth]{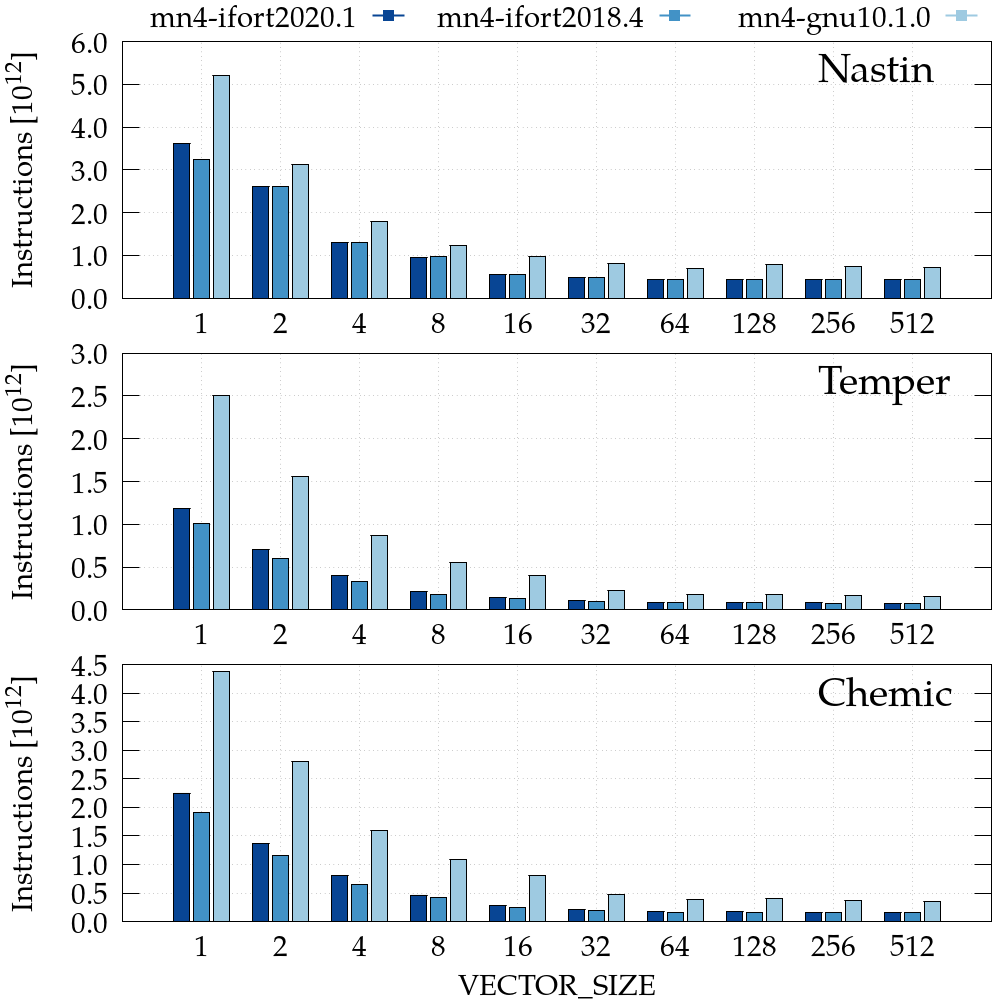}
  \caption{Number of instructions executed in each phase varying the vector size with different compilers}
  \label{figInstructions}
\end{figure}

The next factor that can affect the execution time is the number of instructions.
In this section, we analyze the total number and the type of instructions executed.
The goal is to understand how the \texttt{VECTOR\_SIZE} affects the instruction mix generated by the compilers and the overall performance.
In Figure~\ref{figInstructions}, we show the total number of instructions executed in each phase.
The $x$-axis represents the different values of the  \texttt{VECTOR\_SIZE}, and each series corresponds to the total number of instructions of different compilers ($y$-axis).
We observe that the number of instructions executed when using the GNU compiler is much higher than when using either of the Intel compiler versions.
This difference is more notable in the Temper and Chemic phases than in the Nastin one.
This observation can explain the results shown in subsection~\ref{secTime} when measuring the elapsed time (Figure~\ref{figTimesteps}).

We can see that for some values of \texttt{VECTOR\_SIZE} (1 for Nastin, 1, 2, 4
for Chemic and Temper), the number of instructions executed by the code compiled
with Intel 2020 is higher than the instructions used when compiled with Intel
2018. This also partially explains the difference in time obtained by the two
compilers, as the difference between the two compilers was
 observed for all the phases and all the \texttt{VECTOR\_SIZE} below 8.

For all phases and all compilers, the total number of instructions executed decreases while we increase the \texttt{VECTOR\_SIZE} up to 32. For values of \texttt{VECTOR\_SIZE} greater than 32 there is no impact in the number of instructions and they keep stable when increasing the \texttt{VECTOR\_SIZE}.

\begin{figure}[htbp]
  \centering
  \includegraphics[width=\columnwidth]{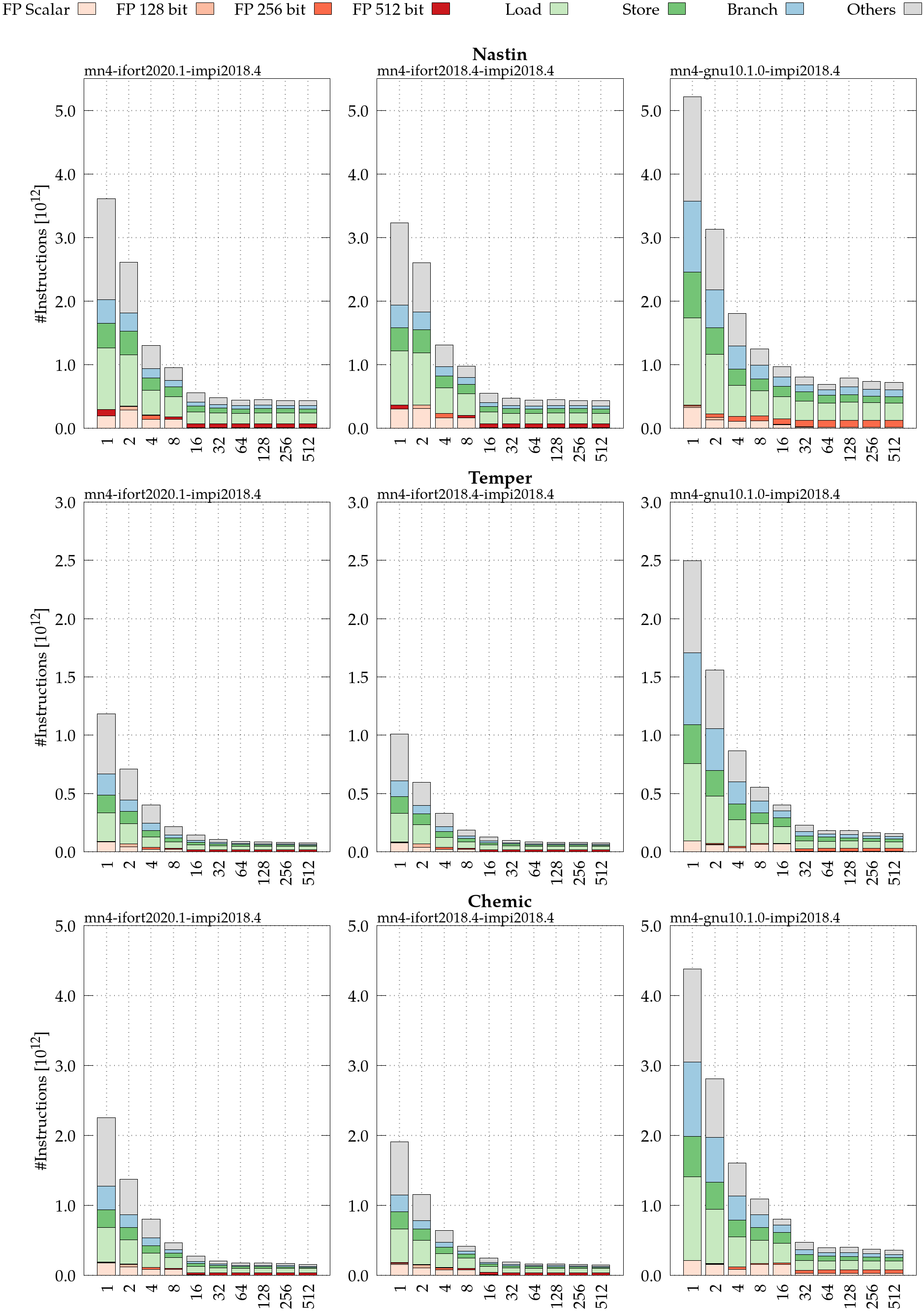}
  \caption{Absolute Instruction Mix for different phases and compilers}
  \label{figInstructionMixAbsolute}
\end{figure}

\begin{figure}[htbp]
  \centering
  \includegraphics[width=\columnwidth]{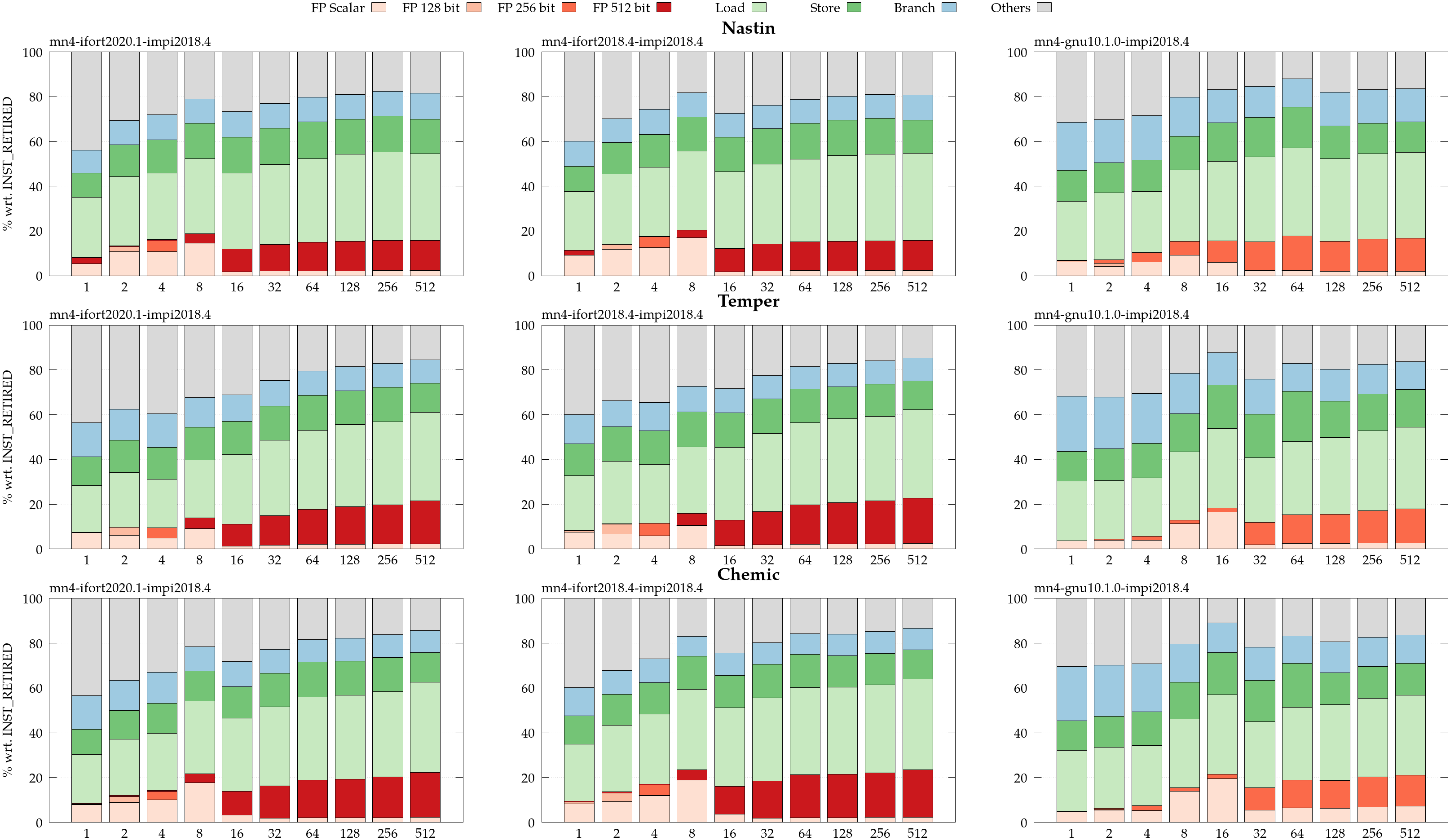}
  \caption{Relative Instruction Mix for different phases and compilers}
  \label{figInstructionMixRelative}
\end{figure}

To understand the cause of the difference in the amount of instruction in each case, we look at the types of instructions executed.
With the available hardware counters in the cluster, we can group the instructions in memory accesses (loads and stores), branches, and floating-point instructions.
Within floating-point instructions we distinguish among instructions operating scalar and vector operands.
Thus, we keep track of floating-point scalar, floating-point 128~bits (SSE), floating-point 256~bits (AVX), and floating-point 512~bits (AVX-512).
We plot this data in two ways: absolute values where the height of the column is equivalent to the total number of instructions (Figure~\ref{figInstructionMixAbsolute}) and relative where all the bars have the same height (Figure~\ref{figInstructionMixRelative}).
Each class of instruction is represented by a color: green for memory accesses, blue for branches and red for floating-point.

In Figure~\ref{figInstructionMixAbsolute}, we show the different plots with the
absolute number of instructions of each type, each column of plots corresponds to
one compiler, and each row of plots corresponds to one phase. To ease the comparison of
phases with different compilers, the plots in the same row share the same
scale in the $y$-axis. In the $x$-axis of each plot we can see the different values of \texttt{VECTOR\_SIZE}.

When comparing the number of instructions executed using the different compilers, we can see that the GNU compiler inserts between 3$\times$ and 5$\times$ more branch instructions, between 1.5$\times$ and 4$\times$ more store instructions, and between 1.5$\times$ and 3.4$\times$ more load instructions.
For \texttt{VECTOR\_SIZE}=1, there is not a very relevant difference in the floating-point instructions: the only observation is that the Intel compiler emits AVX-512 instructions while GNU only generates SSE instructions.
For \texttt{VECTOR\_SIZE}=2, GNU is able to vectorize more than the Intel compiler but it is not enough to overcome the more branch, load, and store instructions executed.

In Nastin, when increasing the \texttt{VECTOR\_SIZE} from 2 to 4 with the Intel compilers, the number of load, store, and branch instructions are divided by~2.
This drastic reduction is not present when increasing \texttt{VECTOR\_SIZE} from 4 to 8, but happens again when increasing from 8 to 16.
As expected, each time that the compiler is able to take advantage of the SIMD units, the absolute number of scalar floating-point instructions executed decreases and also the number of memory accesses and branches decreases proportionally. This explains the instruction reduction when increasing the value of \vs from~2 to~4 and from~8 to~16.

For Temper and Chemic the reduction in the number of load, store, and branch instructions is progressive from \texttt{VECTOR\_SIZE} 2 to 16.

In Figure~\ref{figInstructionMixRelative}, we plot the same data relative to the total number of instructions executed.
In this view, we highlight the differences in the instruction mixes of all cases.
As we increase \texttt{VECTOR\_SIZE} from 1 to 16, we see that the Intel compilers can make use of floating-point vectors of 128 (SSE), 256 (AVX), and 512 bits (AVX-512), respectively.
With \texttt{VECTOR\_SIZE} 16, almost all floating-point operations are vectorized using AVX-512 instructions, \ie FP 512 bits.
The GNU compiler behaves similarly but it is only able to generate AVX instruction (FP 256 bit), hence using half of the vector length compared to the Intel compilers.

From the data shown in this section, we conclude that the difference in performance between the code generated by the GNU and Intel compilers is mainly due to the capacity of the compiler of taking advantage of the CPU vector extension:
we notice in fact that the code transformations in Alya allow to generate floating-point instructions using a vector length of 128~bits (using SSE), 256~bits (using AVX), and 512~bits (using AVX-512). This results in a proportional reduction of memory accesses (load and stores) and branch instructions.

The reduction of the number of instructions also explains the performance improvement achieved when increasing the \texttt{VECTOR\_SIZE} by all compilers in all phases.
In the Intel compilers, the maximum vectorization is achieved with \texttt{VECTOR\_SIZE} 16. For higher values, there is no change in the number of vector instructions. For the GNU compiler, the maximum vectorization is attained with \texttt{VECTOR\_SIZE} 32.  
The performance degradation observed when increasing the \texttt{VECTOR\_SIZE} beyond 32 cannot be explained by the number of instructions nor the instruction mix shown in this section.

%% file: 43-cache.tex
\subsection{IPC and cache reuse analysis}\label{secCache}

In this section, we study $\mathscr{C}$ that is the third factor of the performance equation~\ref{perfEq}, how it changes, how it affects the performance and the causes of its change.

\begin{figure}[htbp]
  \centering
  \includegraphics[width=.95\columnwidth]{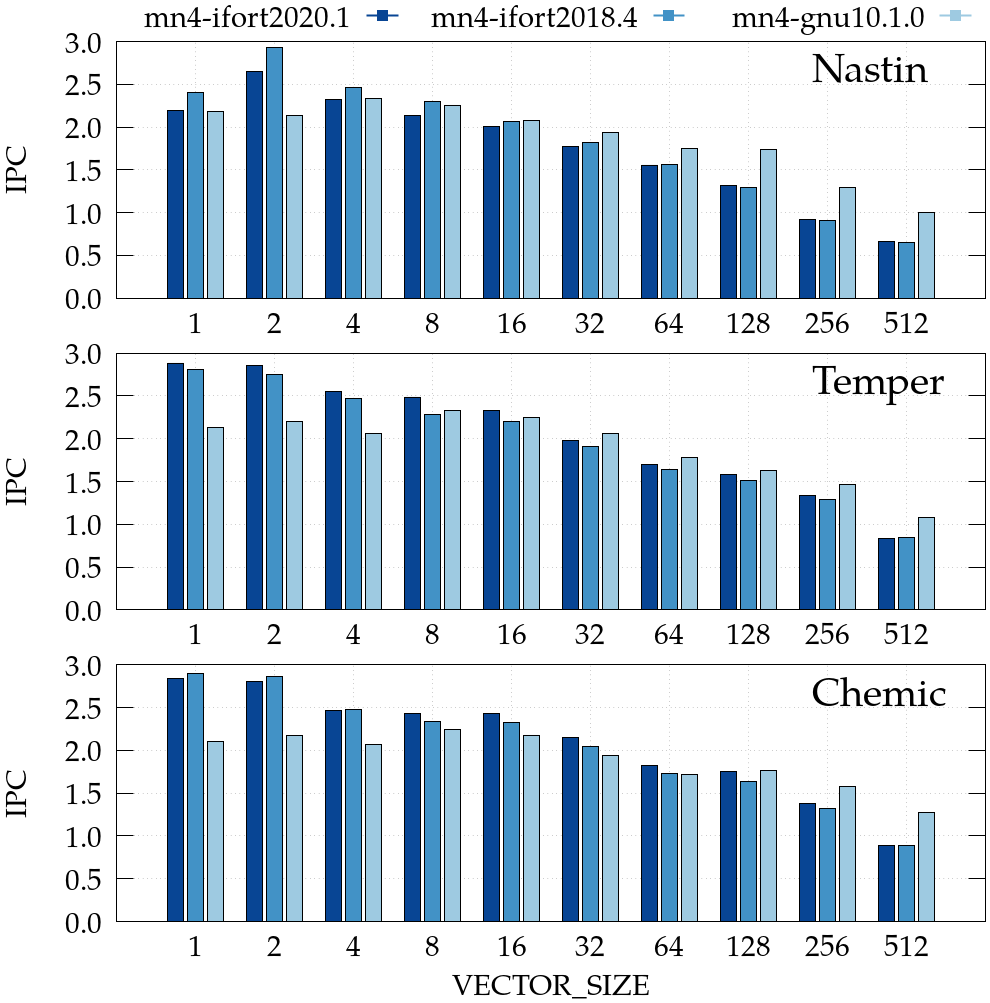}
  \caption{IPC obtained in each phase varying the vector size with different compilers}
  \label{figIPC}
\end{figure}

Figure~\ref{figIPC} shows $\mathscr{C}$ that we call ``Instructions Per Cycle'' (IPC) on the $y$-axis
and the \texttt{VECTOR\_SIZE} in the $x$-axis. Each compiler is represented with a different color. 

We observe that in the Temper and Chemic phases for \texttt{VECTOR\_SIZE} below 8, the GNU compiler obtains less IPC than the Intel compilers.
This means that the performance degradation that we see in Section~\ref{secTime} is explained not only by a higher number of instructions as seen in Section~\ref{secInstrMix} but also by a worse IPC.
For \texttt{VECTOR\_SIZE} above 8 in Nastin and Chemic the GNU compiler obtains a higher IPC than Intel compilers, while in Chemic, this behaviour appears for \texttt{VECTOR\_SIZE} greater than 64.

The difference between the two Intel compilers that is not explained by the number of instructions in Section~\ref{secInstrMix} is explained here. In Nastin Intel 2018 obtains, in fact, a higher IPC than Intel 2020.

In general, increasing the \texttt{VECTOR\_SIZE} decreases the IPC. The pattern is slightly different for Nastin, where the best IPC is obtained with \texttt{VECTOR\_SIZE} 2, while in Temper and Chemic \texttt{VECTOR\_SIZE} 1 and 2 show the same IPC. Increasing the \texttt{VECTOR\_SIZE} above 16 decreases the IPC for all compilers and phases.

\begin{figure}[htbp]
  \centering
  \includegraphics[width=.95\columnwidth]{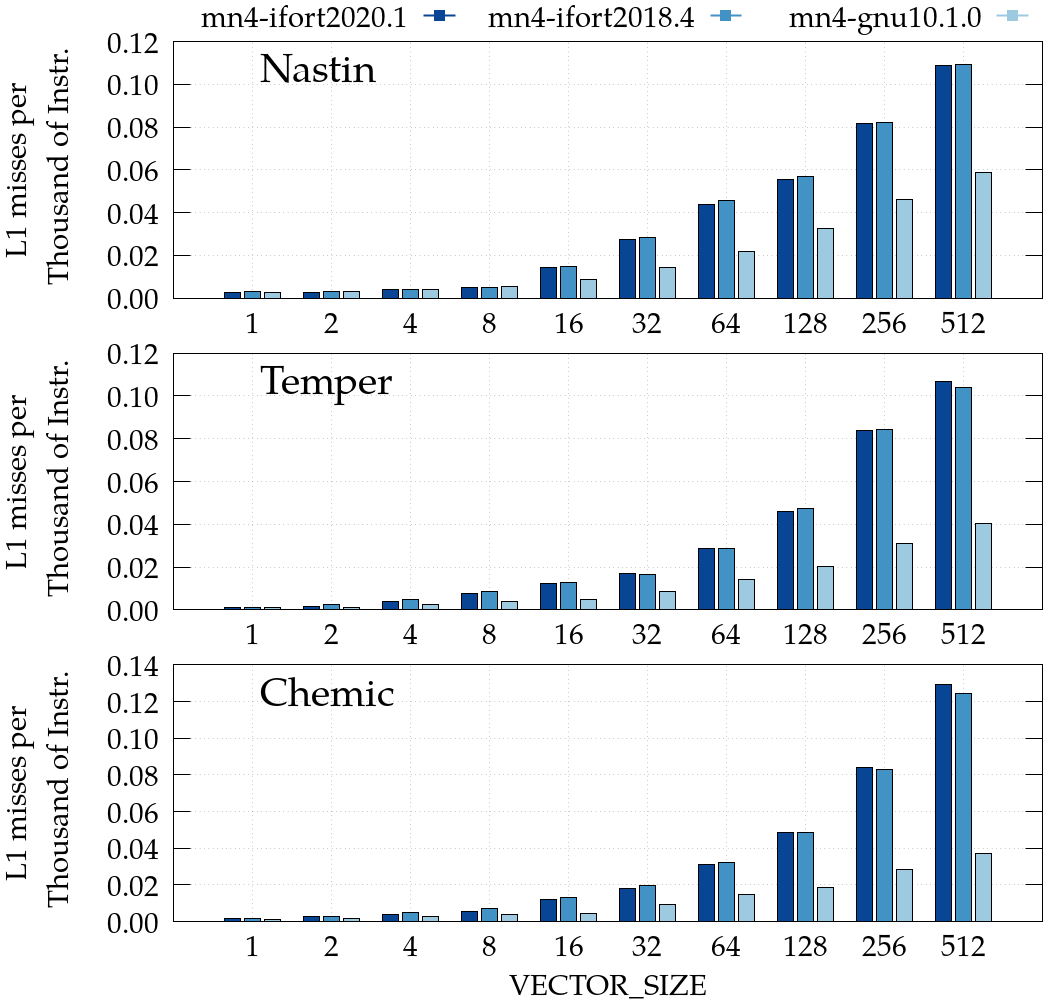}
  \caption{L1 misses per thousand instructions in each phase varying the vector size with different compilers}
  \label{figL1misses}
\end{figure}

To understand the IPC changes observed, we look at the misses in the different levels of cache. Figure~\ref{figL1misses} shows the number of misses issued by the first level cache (L1) for every 1000 instructions.

Comparing the compilers, we observe that the code generated by the GNU compiler obtains fewer L1 misses in all cases.
This can be an effect of the GNU compiler executing more instructions than the Intel compilers.
The two Intel compilers present a very similar miss ratio in L1.

In general, we observe that for \texttt{VECTOR\_SIZE} below 8, there is no relevant variations in the number of L1 misses. On the other hand, for \texttt{VECTOR\_SIZE} above 32, we notice an increment of the number of misses in L1. The changes of \texttt{VECTOR\_SIZE} affect the locality in the L1 cache.

\begin{figure}[htbp]
  \centering
  \includegraphics[width=.95\columnwidth]{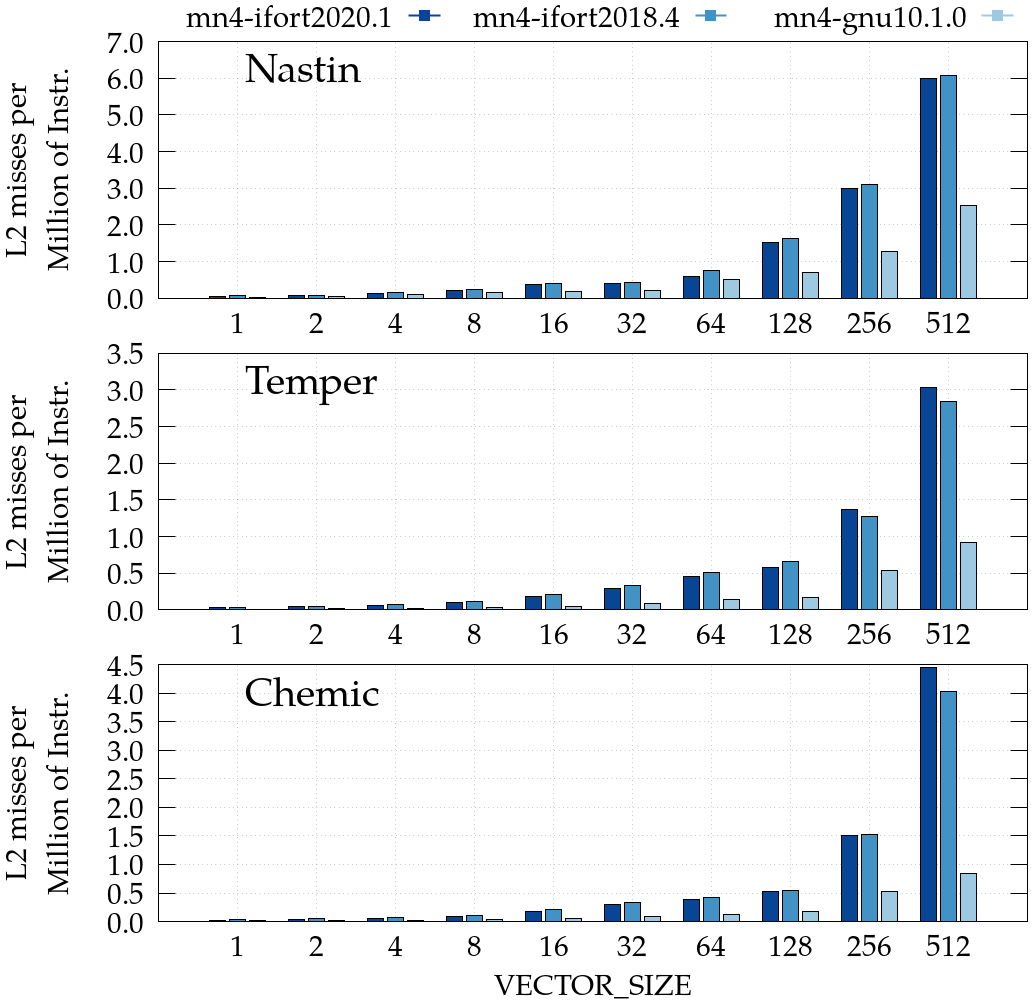}
  \caption{L2 misses per million instructions in each phase varying the vector size with different compilers}
  \label{figL2misses}
\end{figure}

In Figure~\ref{figL2misses}, we can see the analogous chart corresponding to the number of petitions missed in the second level of cache (L2) for every million instructions.
The conclusions are very similar to the ones obtained looking at the L1 misses.
The GNU compiler generates less L2 misses per million of instructions but this metric can be affected by the fact that GNU generates more instructions than the other compilers.
The two Intel compilers show an equivalent number of L2 misses.

For values of \texttt{VECTOR\_SIZE} below 8, the number of L2 misses is very low and does not change when changing the value of \texttt{VECTOR\_SIZE}.
When increasing the \texttt{VECTOR\_SIZE}, we can see an increment in the number of L2 misses. The main difference with the previous chart (L1 misses per thousand instructions, Figure~\ref{figL1misses}) is that the drastic increase starts with a \texttt{VECTOR\_SIZE} above 128. This can indicate that some of the data structures are increasing in size as we increase the \texttt{VECTOR\_SIZE} and do not fit in the caches. This effect can be seen because L1 is saturated first as is smaller and L2 latter.

\begin{figure}[htbp]
  \centering
  \includegraphics[width=.95\columnwidth]{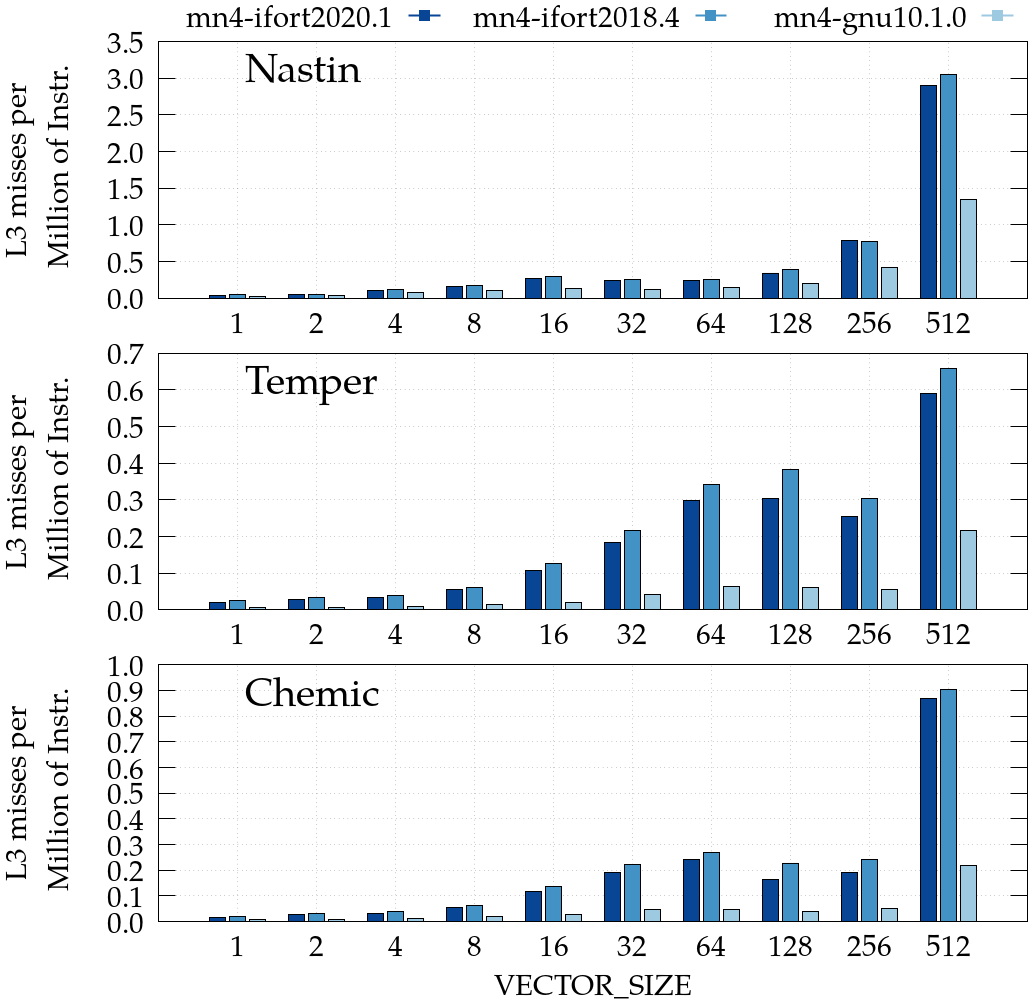}
  \caption{L3 misses per million instructions in each phase varying the vector size with different compilers}
  \label{figL3misses}
\end{figure}

Finally, we look at the number of misses of the last cache level (L3).
We plot them in Figure~\ref{figL3misses}, where the $y$-axis is the total number of misses in L3 per million instructions executed. 
In this case, we can see a different behavior in the different phases.
For Nastin, there is not an important difference in the number of L3 misses for \texttt{VECTOR\_SIZE} below 256 but we see a dramatic increase with \texttt{VECTOR\_SIZE} 512.
This effect follows our previous assumption that some data structure is growing as we increase the \texttt{VECTOR\_SIZE} and fills the different caches as we increase the \texttt{VECTOR\_SIZE}. 

In Temper and Chemic, the L3 miss ratio is relatively stable up to \texttt{VECTOR\_SIZE} 8.
For values of \texttt{VECTOR\_SIZE} between 8 and 64 or 128, the L3 miss ratio increases.
Then, the L3 miss ratio decreases as we increase the \texttt{VECTOR\_SIZE}, and finally, for \texttt{VECTOR\_SIZE} 512, it reaches its maximum value.
With the current information, we cannot explain this effect. However, we can see that this effect is not reflected in the IPC, meaning that the increase in L1 and L2 misses have more impact than the L3 ones in this case.

Analyzing the IPC, we have seen that the better performance observed by the Intel compilers with respect to the GNU one is also explained by the better IPC achieved by the Intel compilers for \texttt{VECTOR\_SIZE} below 8.
When looking into the miss ratio of the different caches, we see that the worse IPC of the GNU compiler is not because of a higher miss ratio.
Therefore, based on the observations in Section~\ref{secInstrMix} about the instruction mix, we can assume that the IPC of the GNU compiler is affected by the higher number of branch instructions.

For \texttt{VECTOR\_SIZE} above 16, the GNU compiler obtains better IPC than the Intel compilers.
This is not explained by the miss ratio of the different levels of cache, so we conclude that it is the effect of the high number of vector instructions executed by the Intel versions.
However, it is important to note that the Intel compilers attain better performance than the GNU compiler in the overall analysis.

Regarding the impact of \texttt{VECTOR\_SIZE}, we conclude that the performance degradation observed for \texttt{VECTOR\_SIZE} greater than 32, which could not be explained in the previous sections, is explained by the decrement of IPC.
Also, the IPC decrement is explained by the increasing miss ratio in the different levels of cache as we rise the value of \texttt{VECTOR\_SIZE}.

%% file: 44-overall-performance.tex
\subsection{Overall performance}\label{secOverallPerformance}

In this section, we evaluate how the proposed changes to the combustion code affect the overall execution.
All results shown in this section are the elapsed time of 10 time integration steps.

\begin{figure}[htbp]
  \centering
  \includegraphics[width=.95\columnwidth]{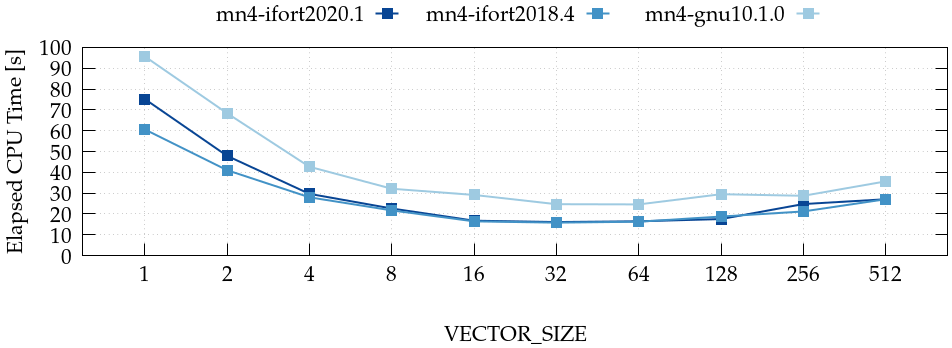}
  \caption{Elapsed time for 10 time integration steps varying the vector size with different compilers}
  \label{figTimestepsGlobal}
\end{figure}

In Figure~\ref{figTimestepsGlobal}, the $y$-axis represents the elapsed time to execute 10 time integration steps of Alya, while the $x$-axis shows the different values used for the \texttt{VECTOR\_SIZE}.
We can conclude that the best \texttt{VECTOR\_SIZE} for the overall execution is 32 for the GNU compiler, while it is around the values of 16, 32, and 64 (within a margin of 5$\%$) for the Intel compilers.
In this plot, we observe that the code generated by the GNU compiler is $1.5\times$ slower than the one generated by the Intel compilers in the best case (with \texttt{VECTOR\_SIZE} 32).
Also, the two Intel compilers have similar performance, except for \texttt{VECTOR\_SIZE} 1 and 2, where Intel 2018 outperforms Intel 2020.

\begin{figure}[htbp]
  \centering
  \includegraphics[width=.95\columnwidth]{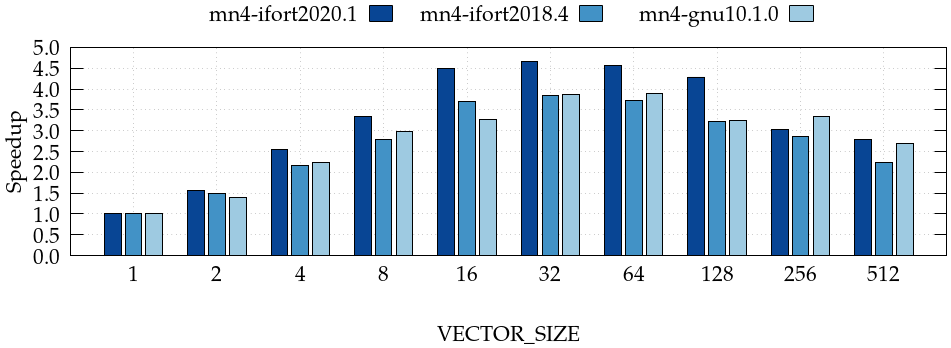}
  \caption{Speedup when increasing \texttt{VECTOR\_SIZE} with different compilers}
  \label{figTimestepsGlobalSpeedup}
\end{figure}

In Figure~\ref{figTimestepsGlobalSpeedup}, we show the speedup of the execution of 10 time integration steps of Alya achieved by increasing the \texttt{VECTOR\_SIZE} with respect to the time spent with \texttt{VECTOR\_SIZE} of 1.
Notice that each compiler uses its own reference for the computation of the speedup.
Compared with \vs 1, we observe that the best \texttt{VECTOR\_SIZE} achieves a speedup of $4.7\times$ with Intel 2020, and $3.9\times$ with Intel 2018 and GNU compilers.
Both Intel 2018 and GNU (the best and worse compilers) obtain the same speedup relative to their base case with \texttt{VECTOR\_SIZE} 1.
This is interesting especially in view of our observations in Section~\ref{secInstrMix} were we show that the GNU compiler is only able to generate AVX instructions (256 bits) while the Intel compiler takes full advantage of the AVX-512 SIMD extension (512 bits).

%% file: 50-conclusiones.tex
\section{Conclusions}\label{secConclusions}

The efficient exploitation of SIMD/vector units is often enforced using non-portable methods: \eg vendors provide optimized libraries that are often tight to the underlying hardware or application developers code part of their code using calls to intrinsics directives.
While both approaches are valid for performance validations on benchmarks or relatively small codes, these approaches could be counterproductive on complex codes with a large community of users and developers that requires to be executed on different HPC clusters (\eg powered by different architectures).
This paper presented a portable method for enabling vectorization within a complex multi-physics code, Alya. We studied the benefits and limitations of our implementation.

Studying the dynamic instruction mix with the help of hardware counters, we have been able to show that 
{\em i)} indeed, our implementation favor vector computation and different compilers are able to exploit it with different degrees of efficiencies;
{\em ii)} pushing to the extreme our implementation proposal (\ie increasing the values of {\tt VECTOR\_SIZE}) affects the data layout in memory, hindering the benefits of data locality in the caches.

The portability of our solution opens the doors to move our code to HPC clusters with different CPU generations or even different architecture, without the need of tuning the code for a new SIMD/vector extension.
While the portability of our solution is a precious added value, we recognize that its efficiency is inherently tight to the ability of the compilers to auto-vectorize our code.
Anyhow, in our study, we show that both compilers, two vendor-specific and the GNU suite, are mature enough on the x86 architecture to enable a high degree of data parallelism using the SIMD units.

Our implementation has been finally evaluated and quantified, showing an overall speed-up respect to the original code ranging from $3.38\times$ up to $4.67\times$ depending on the compiler.

\section{Discussion and Future work}

The portable coding strategy described in the manuscript enables the vectorization independently of the initial conditions of the simulation. The geometry of the elements can negatively affect the performance when dealing with polyhedrons with many faces. Our work was oriented to the more common shapes found in simulations using unstructured grids (tetrahedrons, pentahedrons, and pyramids). 
Our strategy of adding an extra dimension causes an increase in the memory footprint that could reduce the vectorization benefits when dealing with more complex shapes. A future line of work in this direction could be to parametrize and estimate the performance based on the computer architecture and the element shape. Another future work is finding a way to reutilize the SIMD-friendly data structures to exploit GPU devices. A detailed performance analysis would be needed to estimate the achievable performance by following this methodology.

%% file: 90-acks.tex
\section{Acknowledgments}\label{secAcks}

The research activities conducted in this paper have been financed by the CoEC project from the European Union’s Horizon 2020 research and innovation programme under grant agreement N.~952181. Our calculations have been performed on the resources of the Barcelona Supercomputing Center. The authors thankfully acknowledge these institutions.